\documentclass[twocolumn]{aastex62}
\usepackage{graphicx,graphics,amsmath}
\usepackage{natbib}
\usepackage{color}
\usepackage[normalem]{ulem}
\begin{document}

\title{A New Formula for Predicting Solar Cycles}% Force line breaks with \\

\correspondingauthor{Gopal Hazra}
\email{hgopal@iisc.ac.in}

\author{Gopal Hazra}
\affiliation{Dept. of Physics, Indian Institute of Science, Bangalore, India}
\affiliation{School of Space and Environment, Beihang University, Beijing, China}

\author{Arnab Rai Choudhuri}
\affiliation{Dept. of Physics, Indian Institute of Science, Bangalore, India}
%\collaboration{(AAS Journals Data Scientists collaboration)}

%\author{Gopal Hazra$^{1,2}$, Arnab Rai Choudhuri${^1}$}
%\affiliation{$^1$Dept. of Physics, Indian Institute of Science, Bangalore, India\\
%$^2$ Dept. of Space and Environment, Beihang University, Beijing, China}


%\maketitle

\begin{abstract}
A new formula for predicting solar cycles based on the current theoretical understanding of the solar cycle from flux transport dynamo is presented. Two important processes---fluctuations in the Babcock-Leighton mechanism and variations in the meridional circulation, which are believed to be responsible for irregularities of the solar cycle---are constrained by using observational data. We take the polar field near minima of the cycle as a measure of the randomness in the Babcock-Leighton process, and the decay rate near the minima as a consequence of the change in meridional circulation. We couple these two observationally derived quantities into a single formula to predict the amplitude of the future solar cycle. Our new formula suggests that the cycle 25 would be a moderate cycle. Whether this formula for predicting the future solar cycle can be justified theoretically is also discussed using simulations from the flux transport dynamo model.     
\end{abstract}


\section{Introduction}
The sunspot cycle with the approximate period of 11 years is one of the 
most intriguing natural cycles known to mankind.  Solar disturbances,
which become more frequent during the peak of this cycle, control the 
space environment of the Earth and affect our lives in various ways.
Developing a method for predicting the strength of a solar cycle in advance
is of utmost societal importance \citep{Pesnell08, Petrovay10, Chou18}. 
The aim of this paper is to propose a
formula for predicting solar cycles. To apply this formula, we need values
of certain quantities which become available towards the end of the previous
cycle. Once these values are known, it will be possible to use this formula
to predict the forthcoming cycle. We discuss how we arrive at this formula
by analyzing the data of the last few solar cycles.  We also look at the
question whether this formula can be justified on the basis of the flux
transport dynamo model, the theoretical model which has been successful 
in explaining many aspects of the solar cycle.

It has been known that there is a correlation between the polar field of the
Sun during the solar minimum and the strength of the next cycle, allowing
us to use this polar field as a predictor \citep{Svalgaard05,Schatten05}. The
theoretical explanation of this correlation on the basis of the flux
transport dynamo model was provided by \citet{Jiang07}. The poloidal field is generated in the flux transport
dynamo model by the Babcock--Leighton (BL) mechanism from the decay of tilted
bipolar sunspots. Since there is a scatter in the tilt angles of bipolar
sunspots around the average given by Joy's law \citep{Longcope02,Wang15}, the BL
mechanism has an inherent randomness, leading to the unequal production
of the poloidal field in different cycles \citep{KM17}. Since this poloidal field is
brought to the polar region by the meridional circulation to produce the
polar field at the end of the cycle and also diffuses to the bottom of the
convection zone to act as the seed of the next cycle, we have this correlation
between the polar field at the end of a cycle and the strength of the next
cycle. \citet{CCJ07} developed a methodology of incorporating the
randomness of the BL mechanism in the theoretical flux transport dynamo
model and predicted cycle 24 before its onset. Their prediction turned out
to be the first successful prediction of a solar cycle from a theoretical dynamo
model.

When the works mentioned in the last paragraph were being done, it was not yet
realized that there can be another important source of irregularities in
the solar cycle---fluctuations in the meridional circulation (MC). It is the time scale of MC which sets the period of the flux
transport dynamo. A slower MC makes cycles longer. Although we have actual
measurements of MC only during the last few years, durations of past cycles
give an indirect indication how the strength of MC varied with time 
\citep{KarakChou11}. If the diffusion time scale
is shorter than the MC time scale---which is the case if we assume a value
of diffusion based on simple mixing length arguments---then longer cycles
become weaker due to a more prolonged action of diffusion.  
In other words, there would be an anti-correlation between the
duration of the cycle and the cycle strength.  This anti-correlation helps in
explaining some features of observational data such as the Waldmeier effect \citep{KarakChou11}. It
seems that there is a time delay in the effect of MC on the cycle strength.  As
a result, the peak of a cycle depends not on the value of MC at that time, but a
few years earlier \citep{HKBC15}. If the MC was weaker a few years earlier, that would make the
decay rate of the previous cycle smaller.  We actually find a correlation between
the strength of a cycle and the decay rate of the previous cycle \citep{HKBC15}.

{\begin{figure*}[!t]
\begin{center}
\includegraphics[width = 1.0\textwidth]{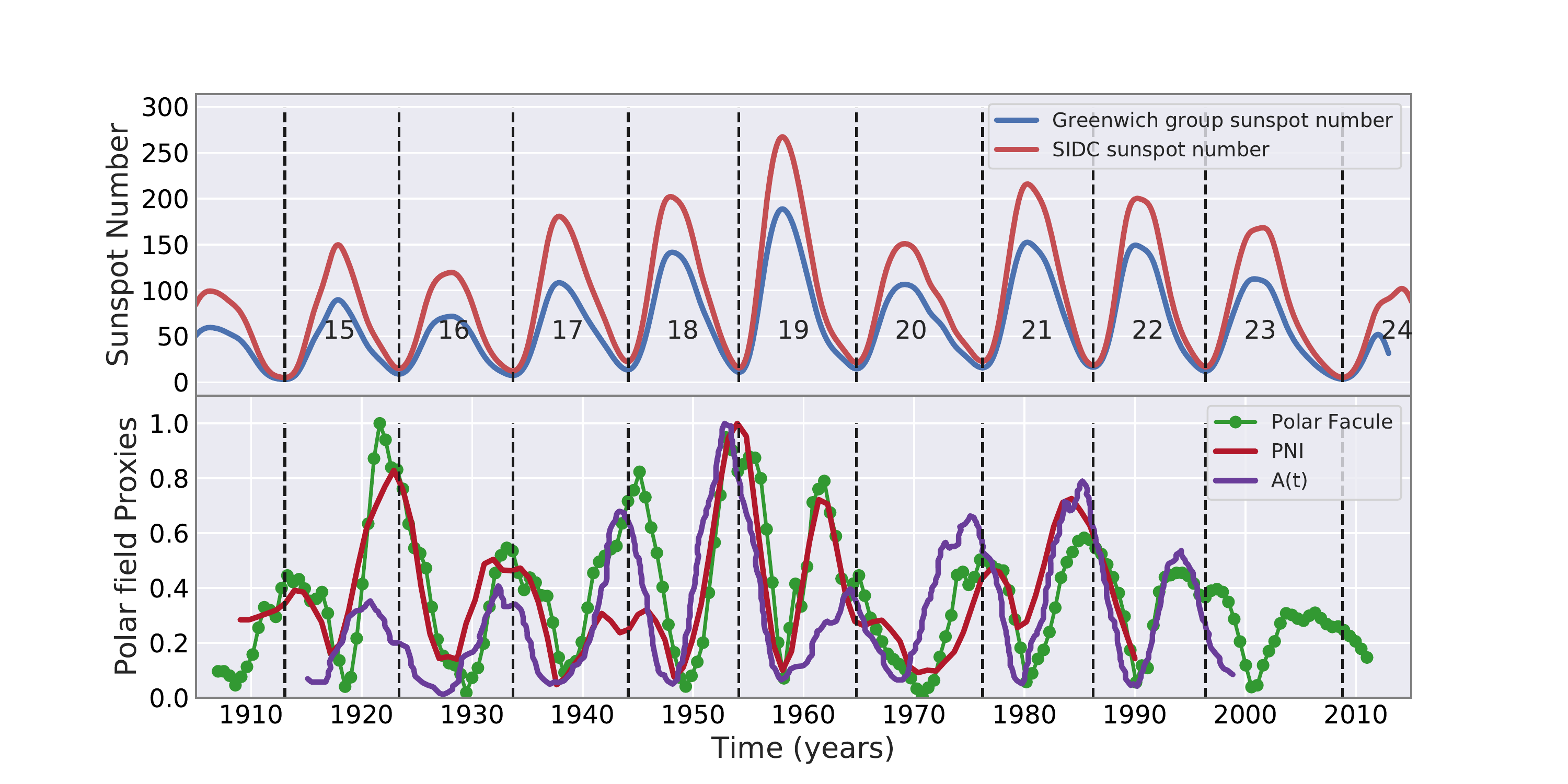}% Here is how to import EPS art
\caption{\label{fig:pfields} The upper panel shows the variation of the sunspot number with time
according to the Greenwich group sunspot number and the SIDC sunspot number (smoothed with FWHM = 2 years). The lower panel shows 
the normalized polar fields from the different available proxies: the polar flux obtained from polar faculae number, the 
$A(t)$ parameter from \citet{Makarov01} and the Polar Network Index from Kodaikanal Observatory. Black dashed vertical lines show the positions of the solar minima.}
\end{center}
\end{figure*}

We conclude that the irregularities of solar cycle are caused by the combined effect
of two factors: (i) randomness in BL mechanism for poloidal field generation, and (ii)
fluctuations in MC. \citet{CK12} developed a theoretical model of the
grand minima of solar cycles by including both of these in their dynamo model.
Now our aim is to develop a method for predicting a future cycle by taking both
these factors into consideration. The polar field $P$ at the sunspot minimum
captures the effect of randomness in the BL mechanism in the previous cycle.  On
the other hand, the decay rate $R$ of the previous cycle provides the information about
MC during the phase which is important for determining the strength of the next
cycle.  So we expect the peak strength $A$ of the cycle to be given by a formula of the
type
\begin{equation}\label{eq:form1}
A \propto P^{\alpha} R^{\beta}.
\end{equation}
Our job now is to check whether the strengths of the past cycles can be matched
by some suitable choices of the indices $\alpha$ and $\beta$. We also look at the question
whether we find any support for such a formula from the simulations of the flux transport
dynamo.

The next Section is devoted to a discussion of the possibility of a formula of type given in Equation~(\ref{eq:form1}) based on observational data. Then
Section~\ref{sec:theory} discusses the support for it in theoretical dynamo simulations.  Whether such a formula
can help us in predicting the upcoming cycle is discussed in Section~\ref{sec:future}. Finally our conclusions
are summarized in the last Section.\\

\begin{figure*}[!htbp]
\begin{center}
\includegraphics[width = 0.95\textwidth]{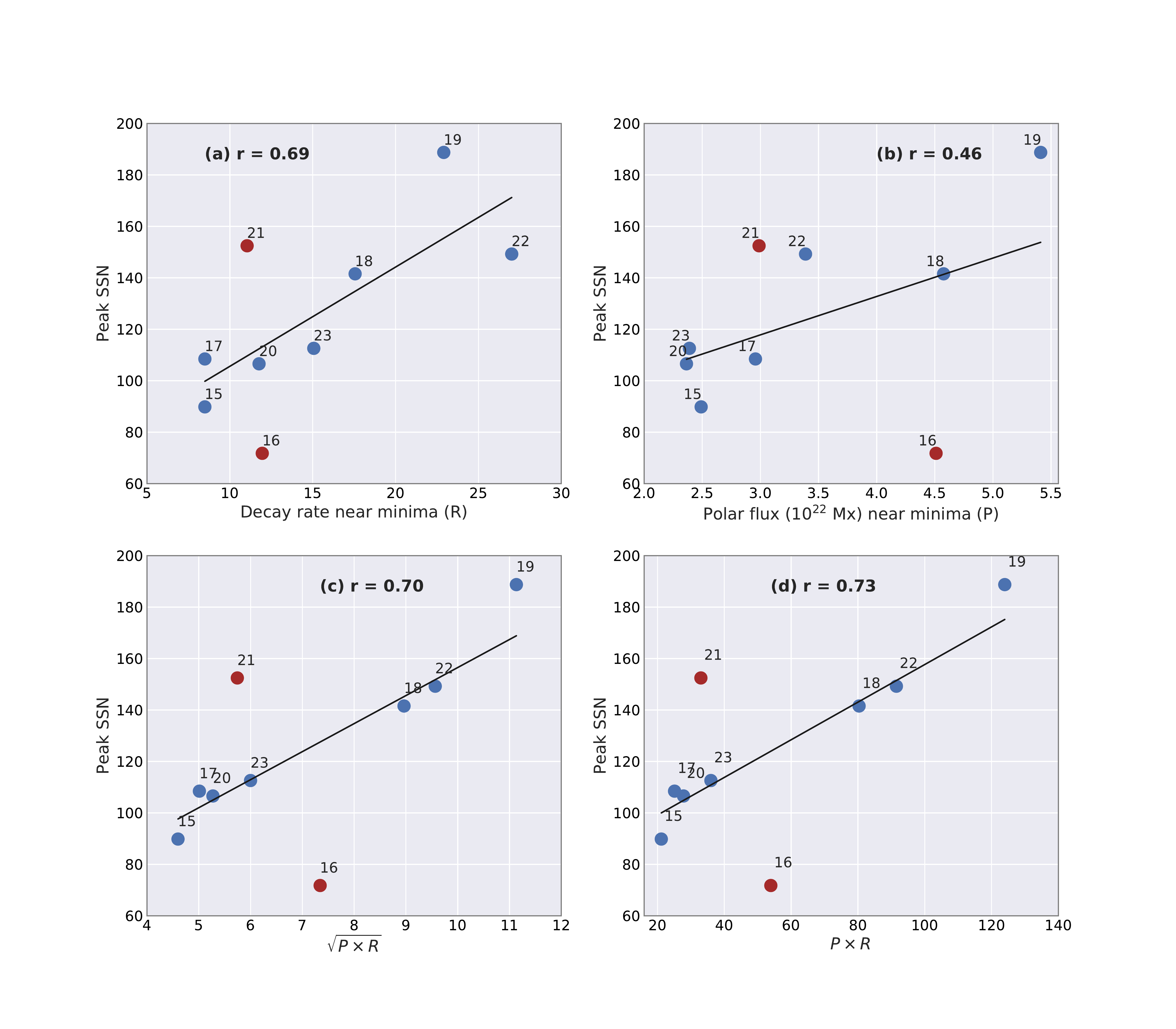}
\caption{\label{fig:greenwich}The correlation of various precursors with the next cycle amplitude is plotted. The
precursors are: (a) the decay rate at the late phase of the cycle, (b) the polar field near minima of the cycle ($P$ is polar
flux in Mx divided by $10^{22}$), (c) the new precursor formula $A= \sqrt{P \times R}$ and (d) $ A = {P \times R}$.}
\end{center}
\end{figure*}

\begin{figure*}[!htbp]
\begin{center}
\includegraphics[width = 0.95\textwidth]{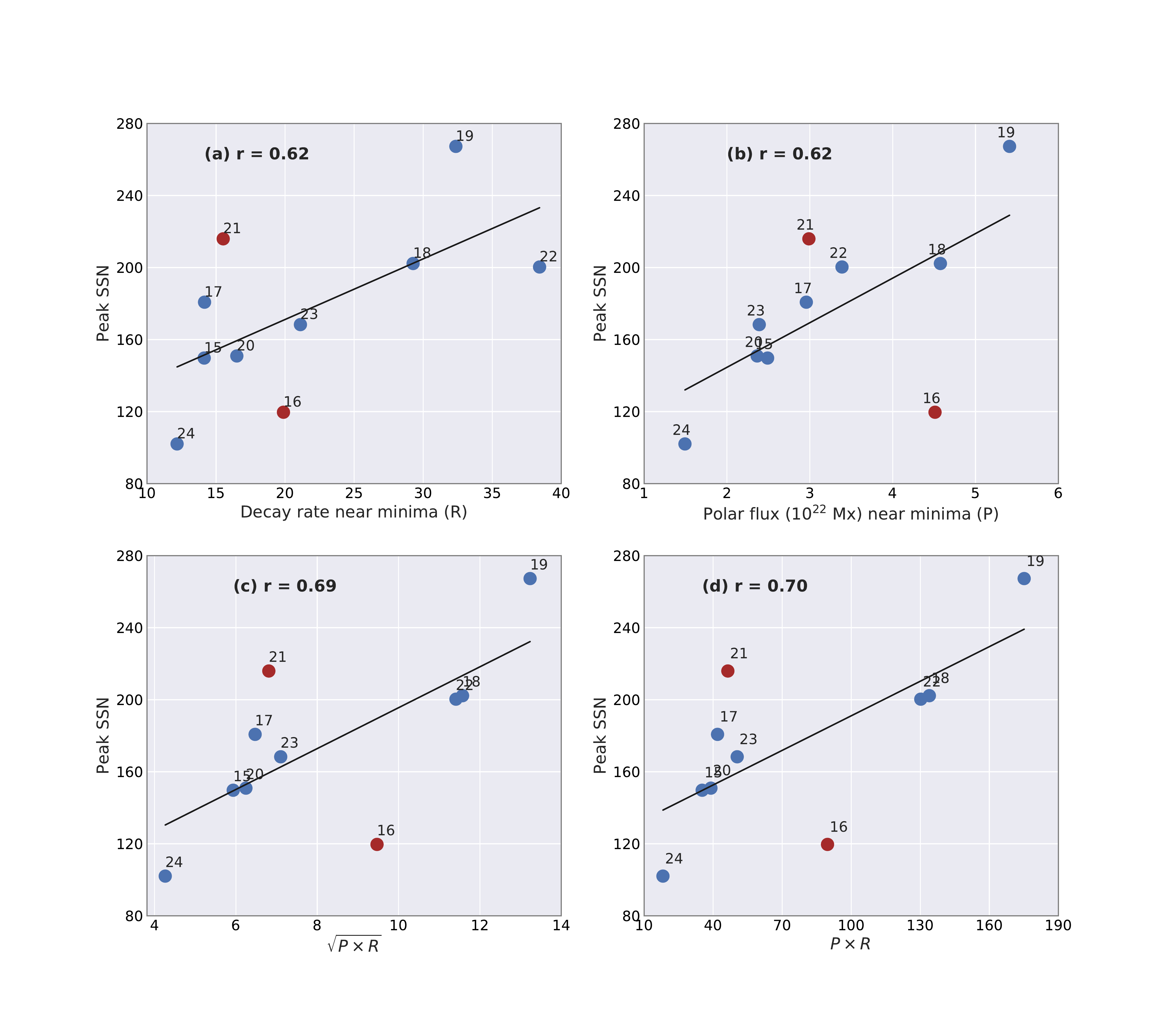}
\caption{\label{fig:sidc} Same as Figure~\ref{fig:greenwich} but for SIDC sunspot number data.}
\end{center}
\end{figure*}

\section{Observational Study}\label{sec:obs}
We have reasonably trustworthy data of sunspot number from at least the beginning
of the 20th century.  We can easily obtain reliable values of the peak sunspot number 
$A$ and the decay rate $R$ for several
previous solar cycles from these data.  To check whether a formula like in Equation~(\ref{eq:form1}) worked
for the past cycles, we need the values of the polar field $P$ during several
sunspot minima.  We have actual regular measurements of the polar field only
from the mid-1970s.  However, there are several proxies which indicate how the polar
field might have evolved at earlier times. The important proxies we consider 
in this paper are:
(i) polar flux obtained from polar faculae number as presented by \citet{Munoz12}; (ii) the parameter $A(t)$ obtained by \citet{Makarov01} from
the position of neutral lines (indicated by filaments) on the solar surface; and
(iii) the Polar Network Index (PNI) developed by \citet{Priyal14} from chromospheric
networks seen in Kodaikanal Ca K spectroheliograms. The bottom panel of Figure~\ref{fig:pfields} plots polar fields 
inferred from these three different proxies, below the sunspot number shown in the
upper panel. For all these three proxies for the polar field, we plot their normalized
values by putting their maximum values in the range equal to 1.
We see that during much of the time the three
proxies give very similar values of the polar magnetic field.  However, during a few solar
minima (such as the minimum before cycle 16), we find that some of the proxies diverge
widely.  We do not know the reason behind this.  Since the polar faculae number happens
to be the most widely studied proxy for the polar field \citep{Sheeley91, Munoz12}, we use the polar field $P$ obtained
from this proxy in our analysis.  We should keep in mind that the value of $P$ obtained
from this proxy may not be very reliable during the times when the different proxies
give very different results. Another possible proxy for the polar field which
we do not include in this paper is the geomagnetic {\it aa } index. \citet{WS09} have studied the correlation of this index at the solar minima with the
strength of the next cycle.

For the sunspot number, we have used two datasets. One, the group sunspot number 
from the Greenwich Observatory 
\footnote{\url{https://solarscience.msfc.nasa.gov/greenwch/spot_num.txt}};
and another one, the calibrated monthly total sunspot number \citep{SIDC14} from WDC-SILSO, Royal Observatory
of Belgium, Brussels 
\footnote{\url{http://www.sidc.be/silso/datafiles}}. 
Both of the datasets are smoothed by
using Gaussian filter with FWHM = 2 years and are shown in the upper panel of Figure~\ref{fig:pfields}. 
Several authors \citep{CS07, Petrovay10,Podladchikova8, Podladchikova17} have suggested possibilities of using some features of sunspot number variations for predicting future cycles.
As we mentioned earlier, the decay rate during the late phase of the
cycle is a good precursor to predict the next cycle. We now calculate the decay rate during the late phase
of the cycle following the procedure used by \citet{HKBC15}, in which 
the decay rate is taken as the slope between two points with a separation of 1 year with the second point 1 year before the minimum at the end of the cycle.
The decay rate defined in this way is found to have a good correlation with
the next cycle, which is not the case if the decay rate is defined in other ways
\citep{HKBC15}. Presumably, the decay rate calculated in one particular way captures
the information about the strength of MC at a relevant phase of the cycle
that affects the next cycle. 

We first present the results obtained by
using the Greenwich sunspot number. Then we shall present results based on SIDC data.
Figure~\ref{fig:greenwich} presents the results we get by using the Greenwich sunspot number. Figure~\ref{fig:greenwich}(a) shows the correlation 
between the decay rate calculated during the late phase of the cycle and the peak amplitude of the next cycle. The correlation 
coefficient $r = 0.69$ is less than $r = 0.83$ that \citet{HKBC15} obtained by considering 23 cycles with same Gaussian
smoothing with FWHM = 2 years (see table 1 of \citet{HKBC15}). In all of our calculations, we have considered the data 
from cycle 15 up to the present time, as the polar faculae data are not available before cycle 15. 

The different campaigns of the
polar faculae data  are calibrated with the direct measurement of the polar fields from Wilcox solar observatory
from the mid-1970s. We use
the calibrated long-term polar flux dataset for the polar fields presented by \citet{Munoz12}. We calculate
the polar field near the solar minima by averaging over a span of one year before the minimum and one year after the minimum.    
The correlation between polar field near minima and the next cycle amplitude 
is shown in Figure~\ref{fig:greenwich}(b). For the polar field $P$, we have used the polar flux (in Mx) which is obtained by
multiplying the normalized polar field shown in Figure~\ref{fig:pfields} by a factor of $6.25 \times 10^{22}$ following the calibration
given by \citet{Munoz12}. We see that
the correlation ($r = 0.46$) is rather poor---mainly due to the reason that cycles 16 and 21 
show large scatters. It may be noted that Figure~2(a) of \citet{Munoz13b} presented
essentially the same correlation ($r = 0.69$) that we present in our Figure~2(b), except some
differences (they use sunspot area data, whereas we use sunspot number data). We
point out that we get a lower correlation than \citet{Munoz13b}. We shall 
discuss below that the correlation improves considerably on using SIDC sunspot
number data (Figure~3(b)) rather than the Greenwich sunspot number data. 
%, but does
%not become as high as what \citet{Munoz13b} find.} 

Finally, to check whether $A= \sqrt{P \times R}$ and $ A = {P \times R}$ 
can be used as good precursors for
predicting the next cycle, Figure~ \ref{fig:greenwich}(c) and \ref{fig:greenwich}(d) show the correlations of the peak of the next cycle
with them. We find that among all possible values of $\alpha$ and $\beta$ appearing in (1), the combination of $\alpha = \beta= 0.5$ 
[case (i)] and $\alpha = \beta =1$ [case (ii)] give the highest correlations with the amplitude of the next cycle. The correlation 
coefficients for case (i) is 0.70 (null hypothesis rejected with probability 96.3$\%$) and for case (ii) is 0.73 (null hypothesis 
rejected with probability 97.3$\%$). These correlation coefficients are significantly higher than the correlation coefficient in the case of 
polar fields alone and slightly higher than the correlation coefficient in the case of decay rates alone. 

It should be clear from Figure~\ref{fig:greenwich} that the points corresponding to cycles 16 and 21 in all the plots make correlation
coefficients lower than what they would otherwise be.  These points are indicated in brown color. 
Interestingly, we see in Figure~\ref{fig:pfields} that the various proxies of the polar field preceding these cycles
did not match each other well.  We have no explanation for this.  However, this raises the question whether polar faculae
counts at these times were good indicators for the polar field.  In particular, the cycle 16 was the weakest cycle in the 
century. Since, according to the flux transport dynamo model with reasonably high diffusivity \citep{Jiang07}, the polar field strengths during the minima preceding 
the cycles are expected to be correlated with the strengths of the cycles, we expect the polar field before the cycle 16 would be weak. 
But in the Polar faculae count and PNI, the polar field is comparable to the polar field preceding the strongest cycle 19. However, 
the polar field from $A(t)$ index of \citet{Makarov01} before cycle 16 is weak as expected from theory. We also point out that,
during the minima preceding cycles 20 and 21, there was a dearth of spectroheliogram plates 
and some error may be introduced in the PNI count  
during these times \citep{Priyal14}. Had we used $A(t)$ as the proxy of the polar field rather than 
the polar faculae count, then the points corresponding
to cycles 16 and 21 in Figure~\ref{fig:greenwich} would have considerably less scatter. In Figure~2 of \citet{WS09} also, we see in a plot
of {\it aa} index at minima against next cycles that the points for cycles
16 and 21 are not so far from the best-fit straight line.
Since we are unsure of the polar field during the minima
preceding cycles 16 and 21, we quote the correlation coefficients for our newly 
proposed precursors without the points 16 and 21. 
The correlation coefficient for case (i) with square root dependence on polar field and decay rate is 0.97 with null hypothesis rejected with probability 99.9$\%$. For case (ii), the correlation coefficient without cycle 16 and 21 is 0.98 (99.9$\%$). Having these high values of correlation coefficients, we argue that 
the newly proposed precursor formulae hold a promise to predict the future solar cycle based on the polar field measurement $P$ preceding the cycle 
and the decay rate $R$ at late phase of the previous cycle. These two quantities represent the two physical causes behind irregularities in the solar cycles, 
namely fluctuations in the BL mechanism and fluctuations in the meridional circulation (MC) respectively. We also point out that, without the cycles 16 and 21, 
the correlation coefficient for polar field alone with the next cycle amplitude is around 0.90 (99.5$\%$). This is significantly higher than what we get
when these cycles are included, but less than the the correlation coefficients obtained with the newly suggested precursors. Therefore, the
precursors we are suggesting would be very helpful in predicting a future solar cycle with better accuracy.    

Next, we repeat the same exercise with the newly calibrated international sunspot numbers from SIDC, Royal Observatory of Belgium. The results are 
presented in Figure~\ref{fig:sidc}. Each panel in Figure~\ref{fig:sidc} is similar to the Figure~\ref{fig:greenwich}. It is clear from 
Figure~\ref{fig:sidc} that the newly proposed precursors are more highly correlated with the peak amplitude of the cycle
compared to either the decay rate of the previous cycle alone or polar field during the preceding minima alone. 
We find that the correlation of decay rate with the next cycle amplitude is 0.62 (94.5$\%$) and the correlation of polar fields near minima with the 
next cycle is also 0.62 (94.5$\%$). On the other hand, the newly proposed precursors (Figures~\ref{fig:sidc}(c) and (d)) have correlation coefficients 
of 0.69 (97.4$\%$) for case (i) and 0.70 (97.6$\%$) for case (ii), which are higher than the correlation coefficients found from the polar field alone (Figure~\ref{fig:sidc}(a)) and the decay rate alone (Figure~\ref{fig:sidc}(b)). The correlation coefficients do get improved if we exclude 
points 16 and 21. However, when we use SIDC data, we find that the correlation coefficient obtained from the polar field alone becomes quite high and
the correlation coefficients obtained with the new precursor formulae (for each case (i) and case (ii)) become slightly less. This shows the hazard of doing
statistical analysis with very few data points. 

\begin{figure*}[!htbp]
\begin{center}
\includegraphics[width = 0.95\textwidth]{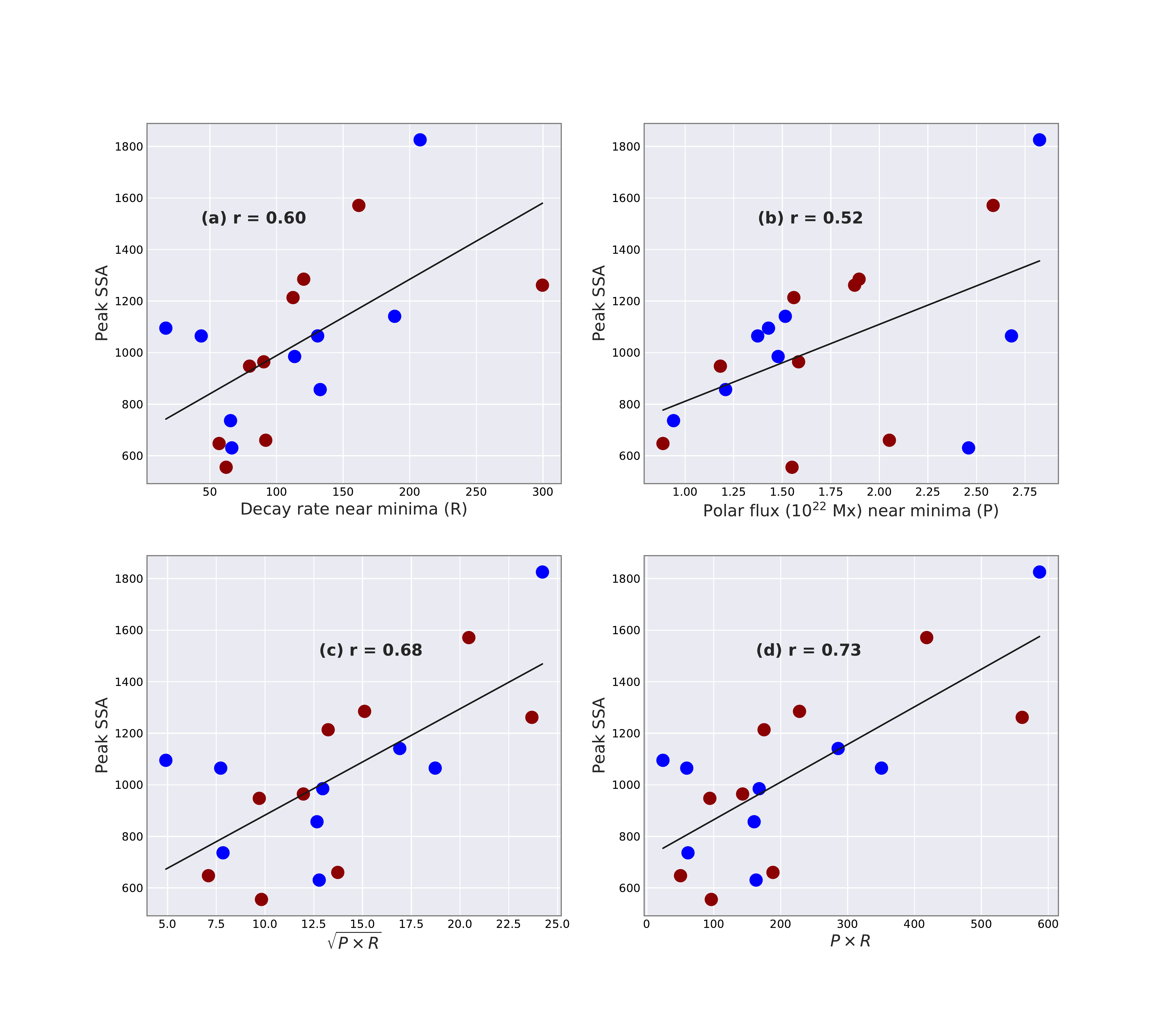}
\caption{\label{fig:hemispheres} Same as Figure~\ref{fig:greenwich} but for data combining northern and southern hemispheres, which are treated separately. Blue and red data points are from northern and southern hemispheres respectively.}
\end{center}
\end{figure*}

We also checked whether the trends we found exist if we analyze the data
of two hemispheres separately.  Since the data for two hemispheres are available
separately only from the Greenwich Observatory, we use these data to calculate
strength of the cycle and the decay rate $R$---by using the same method we
are using, but now doing the analysis for the two hemispheres separately. 
Figure~\ref{fig:hemispheres}, which is similar to Figure~\ref{fig:greenwich} and Figure~\ref{fig:sidc}, shows results of this analysis by indicating data points
corresponding to northern and southern hemispheres in blue and red respectively.
Table~\ref{tab:hemi} shows the correlation coefficients we would get if we treat the data
of the two hemispheres separately instead of combining them together, as we
have done in Figure~\ref{fig:hemispheres}. Even in this analysis, we always find the trend that $A= \sqrt{P \times R}$ and $ A = {P \times R}$ are correlated better
with the next cycle than $P$ or $R$, giving an indication that this trend may
not be an artifact of a small data set.

\begin{table}[]
\centering
\caption{Correlation coefficients of different precursors (e.g., Polar Field (P), Decay Rate near late phase (R), and newly defined precursors) with the next cycle amplitude considering data from Northern Hemisphere (NH), Southern Hemisphere (SH) and combining both. \label{tab:hemi}}
\begin{tabular}{c|c|c|c|c}
\hline
\hline
Data & \multicolumn{4}{c}{Correlation of next cycle amplitude with}\\
\cline{2-5}
 & $P$ & $R$ & $\sqrt{P \times R}$ & $P \times R$\\
\hline
NH & 0.47 (79.7\%) & 0.60 (91.1\%) & 0.64 (93.4\%) & 0.76 (98.3\%) \\
SH & 0.61 (92.2\%) & 0.63 (93.0\%) & 0.75 (98.1\%) & 0.71 (96.8\%) \\
All & 0.52 (97.4\%) & 0.60 (99.2\%) & 0.68 (99.8\%) & 0.73 (99.9\%) \\
\hline
\hline
\end{tabular}
\end{table}
Although our analysis is severely restricted by the very limited number of data points and the values of
the polar field are uncertain in some cases even among these few data points, we still find tantalizing hints that the new precursor formulae we are suggesting
are better for predicting a future cycle than the polar field at the minimum alone or the decay rate of the previous cycle alone. We have to wait for a 
few more cycles (at least for half a century) before one can draw firmer conclusions.

\begin{figure*}[!htbp]
\begin{center}
\includegraphics[width = 0.95\textwidth]{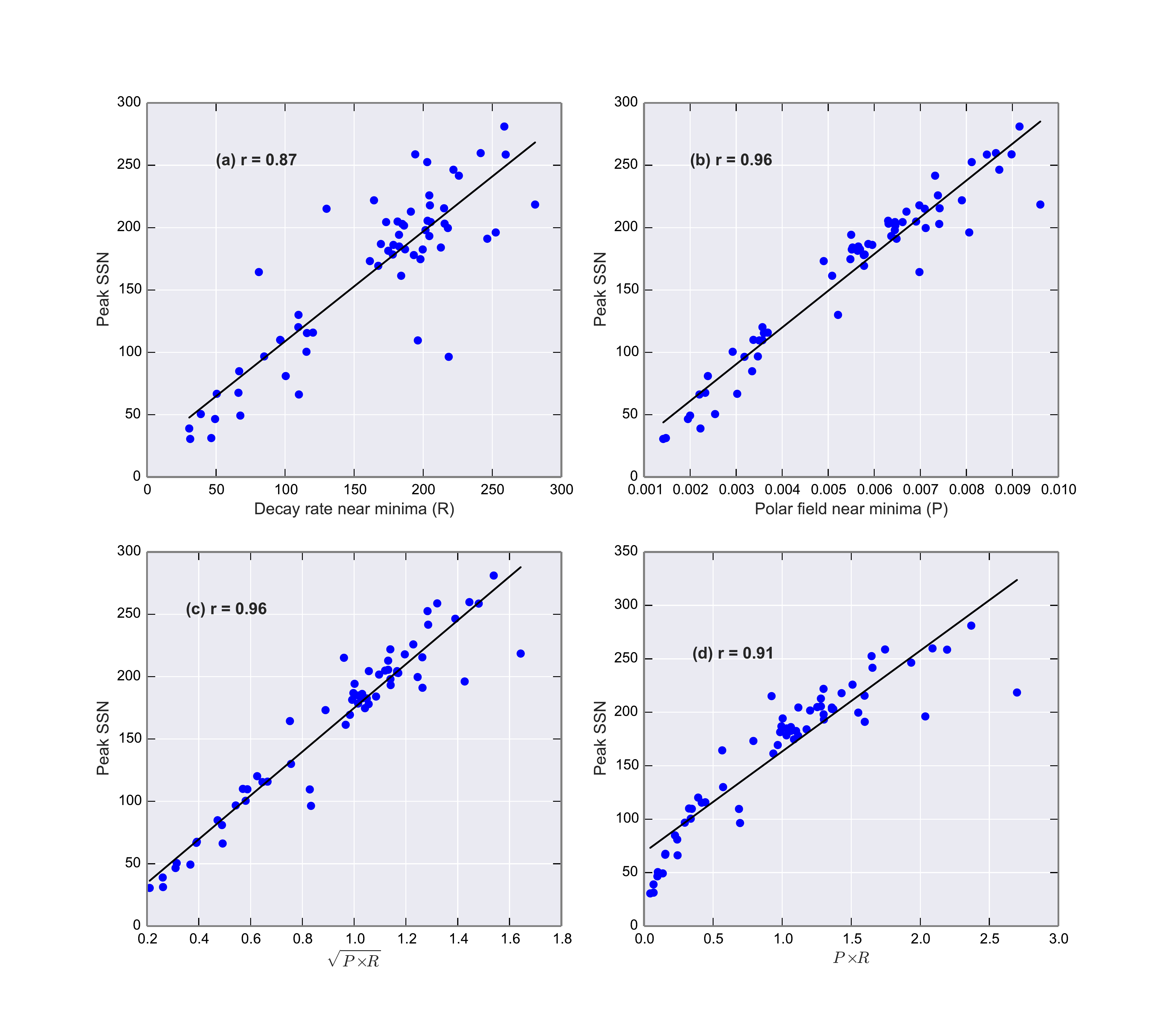}
\caption{\label{fig:theory} Same as Figure~\ref{fig:greenwich} but for theoretically produced irregular solar cycles.}
\end{center}
\end{figure*}

\section{Theoretical Interpretation}\label{sec:theory}           
As explained in the previous Section~\ref{sec:obs}, our observational study motivated us to introduce new precursors 
for predicting the future solar cycle. In this section, we discuss whether we can provide any justification for 
these new precursors from a theoretical flux transport dynamo model. Presently, the flux transport dynamo model is 
the most promising model to explain the various features of the solar cycle and its irregularities. 

The cyclic oscillation between the poloidal field and the toroidal field produces the solar cycle in this model \citep{CSD95, 
Durney95,CD99,CNC04}. The differential rotation stretches the poloidal field to generate the toroidal field. Then the
toroidal field rises up to the photosphere due to magnetic buoyancy and creates bipolar magnetic regions which appear
with tilts produced by the Coriolis force \citep{Dsilva93}. The 
decay of tilted bipolar magnetic regions due to turbulent diffusion produces the poloidal field via the Babcock-Leighton 
mechanism \citep{Bab61, Leighton69}. For a detailed explanation of the model, please see the reviews 
by \citet{Chou11}, \citet{Charbonneau14} and \citet{Karakreview14}.
The irregularities in the solar cycle mainly arise because of the inherent randomness in the mechanism for generating the poloidal field, as first pointed out by \citet{Chou92} and then analyzed by many authors \citep{CD2000, KarakChou11, HKBC15} who successfully reproduced many observed irregularities in the solar cycle. But some of the irregular properties (e.g., the Waldmeier Effect, the correlation between decay rate and the amplitude of the next cycle) are not reproduced using only the fluctuations in the Babcock-Leighton mechanism. \citet{Karak10} pointed out that the fluctuations in the meridional circulation can be another source of irregularities in the solar cycle. By including fluctuations in the meridional circulation, \citet{KarakChou11} successfully reproduced the Waldmeier effect and \citet{HKBC15} reproduced the correlation of the decay rate with the next cycle amplitude. 

We now present some theoretical results obtained by introducing fluctuations in both the Babcock-Leighton process 
and the meridional circulation in our dynamo model. We solve the time evolution equations of the poloidal and toroidal fields as given below.
\begin{equation}\label{eq:Aeq}
\frac{\partial A}{\partial t} + \frac{1}{s}({\bf v}.\nabla)(s A)
= \eta_{p} \left( \nabla^2 - \frac{1}{s^2} \right) A + S(r, \theta, t),
\end{equation}
\begin{eqnarray}
\label{eq:Beq}
\frac{\partial B}{\partial t}
+ \frac{1}{r} \left[ \frac{\partial}{\partial r}
(r v_r B) + \frac{\partial}{\partial \theta}(v_{\theta} B) \right]
= \eta_{t} \left( \nabla^2 - \frac{1}{s^2} \right) B ~~\\ \nonumber
  + s({\bf B}_p.{\bf \nabla})\Omega~~~~ \\ \nonumber
  + \frac{1}{r}\frac{d\eta_t}{dr}\frac{\partial{(rB)}}{\partial{r}}
\end{eqnarray}
where $A {\bf e}_{\phi}$ is the magnetic vector potential corresponding to the poloidal field, $B$ is the toroidal component and $s = r \sin \theta$. The
diffusion coefficients ${\eta_t}$ and ${\eta_p}$ correspond
to poloidal and toroidal components of magnetic field respectively. The source term $S(r,\theta)$ takes care of magnetic buoyancy and the Babcock-Leighton mechanism. We have considered a local $\alpha$-parameterization of magnetic buoyancy \citep{CH16}. 
The parameters for the flux transport models are chosen 
the same as in Section~4.3 of \citet{HKBC15}.
If the toroidal field near the bottom of convection zone is stronger than a critical value $B_c$, some amount of toroidal field is removed from there and put it to the surface layers, which produces the poloidal field via Babcock-Leighton mechanism. This introduces a nonlinearity in the problem which limits the dynamo growth.

To produce irregularities in the cycles, 100$\%$ fluctuation is introduced in the Babcock-Leighton $\alpha$ with coherence time of one month, whereas a 30$\%$ fluctuation with 30 years of coherence time is introduced in the meridional flow. 
This gives us reasonably irregular cycles comparable to observed solar cycles.  The decay rate during the late phase of the cycle is calculated from
the output of the theoretical simulation and its correlation with the amplitude of the next cycle is shown in 
Figure~\ref{fig:theory}(a). This is the same as the results presented in \citet{HKBC15}. We have calculated the peak 
polar field near the minima of the cycles.  The correlation between this polar field near minima and the next cycle 
amplitude is shown in Figure~\ref{fig:theory}(b). The two bottom panels of Figure~\ref{fig:theory} show how well
the new precursors introduced by us on the basis of the observational data are correlated with the amplitude
of the next cycle.  It may be noted that the theoretical correlation between the polar field at the minima and the
peak of the next cycle, as shown in Figure~\ref{fig:theory}(b), is already very high---considerably higher than what is
seen in the observational data (see Figures~\ref{fig:greenwich}(b) and \ref{fig:sidc}(b)). In such a situation, it is somewhat difficult to ascertain
whether the new precursors give even better correlations. We made several independent runs of our code and found
that the correlations computed in different runs are often slightly different, although they have the same
statistical nature. In Figure~\ref{fig:theory} we
have presented results from one run in which the both the precursors $\sqrt{(P \times R)}$ (case (i)) and
${(P \times R)}$ (case (ii)) give better correlations with the next cycle amplitude than $R$, but about
the same as $P$. In fact, the correlation coefficients for the precursors are marginally lower than that
for $P$. Since $P$ alone gives such a good correlation in our theoretical model, precursors which combine
it with the less strongly correlated quantity $R$ tend to have slightly less correlations.

To verify theoretically whether our proposed precursors are really better for predicting a future cycle than either $P$ or $R$,
we need a theoretical dynamo model which faithfully reproduces all the different features of the solar cycle. The model
presented in Section~4.3 of \citet{HKBC15} reproduced most of the features of the solar cycle.  However, we now realize
that it produces a tighter correlation between the polar field $P$ at the minima and the next cycle amplitude compared to
what is observed.  As pointed out by \citet{Jiang07}, this correlation becomes better on increasing the value of
the turbulent diffusion.  So we need to do calculations with a model having a lower value of diffusion in order to increase
the scatter in the correlation between $P$ and the next cycle.  On the other hand, \citet{CNC04} showed that
the dynamo solution tends to become quadrupolar (contrary to what is observed) on decreasing the value of diffusion.  It
is, therefore, needed to construct a theoretical model which prefers dipolar parity but gives more scatter in the correlation
between $P$ and the next cycle.  On the basis of several trial runs using different combinations of parameters, so far we
have not been able to come up with a theoretical model which has this property.  So we are right now unable to give very
strong arguments on the basis of a theoretical dynamo model that the precursors suggested by us are better in predicting
a future cycle than $P$ and $R$. Based on the theoretical calculations we have presented, we can certainly say that our 
precursors are very good for predicting the next cycle, although the correlation of $P$ with the next cycle is already so
strong in the theoretical model that it is not clear whether the precursors correspond to significant improvements. 

Another issue we should mention is that the effect of the variations of meridional flow on the polar field can be different in models handling the
BL mechanism differently \citep{Munoz10}. Although treating the BL mechanism
through an $\alpha$-parameterization led to reasonably realistic theoretical
models matching observations, we should keep in mind that this is a gross
over-simplification and the results should be interpreted with caution
when there are variations in the meridional circulation.

In summary, we emphasize that, given the many uncertainties in both the 
theoretical model and the observational data, the comparison between
the two should be taken to be of
illustrative nature only and should not be taken too seriously as a
suggested real physical interpretation.

\section{Future Cycle Prediction}\label{sec:future}
As we believe that the two precursors we have suggested (case(i) and case(ii)) are particularly well suited to predict future cycles, we now write down appropriate formulae based on these precursors which can be readily used for predicting future cycles.  
Since the sunspot database from SIDC is the best calibrated and most trustworthy sunspot number database, we consider the best straight line fits of the data points for the two cases of the SIDC database only (Figures~\ref{fig:sidc}(c) and \ref{fig:sidc}(d)).    
We perform the least square fitting for both the cases, i.e. case (i) with $\alpha = \beta = 0.5$ (Figure \ref{fig:sidc}(c)) and case (ii) with $\alpha = \beta =1.0$ (Figure~\ref{fig:sidc}(d)), to arrive at formulae from which the amplitude of the next solar cycle can be calculated, after knowing the values of the precursors $P$ and $R$ around the minimum before the cycle. The formula representing the case (i) ($\alpha = \beta =0.5$) is
\begin{equation}\label{formulla1}
A =  11.35 \times \sqrt{P \times R} + 82.03,
\end{equation}
whereas for case (ii) (with $\alpha = \beta =1$) the formula is
\begin{equation}\label{formulla2}
A =  0.64 \times (P \times R) + 127.07.
\end{equation}  
Note that the best fit straight lines in Figures~\ref{fig:sidc}(c) and \ref{fig:sidc}(d) do not pass through the origin.  In other words, a future cycle is never predicted to have zero strength for any combination of positive values of $P$ and $R$. It is thus clear that the formulae we have arrived at 
cannot handle the situation of a grand minimum.  Presumably, these formulae would
give good results when the various parameters lie within a reasonable range of values. When we know the appropriate values of the polar field $P$ and decay rate $R$
(of the previous cycle) at the time of a solar minimum, we can use these formulae for predicting the next cycle.   

Finally, we calculate the peak sunspot number of cycle 25 based on our newly obtained precursor formulae given in Equations~(\ref{formulla1}) and (\ref{formulla2}). 
As these formulae need $P$ and $R$ values, we calculate them individually. 
Ideally, in conformity with what we are doing for the other cycles, we
should calculate $P$ by averaging over 2 years around the minimum and should
calculate $R$ from the slope of sunspot number over a year ending 1 year before
the minimum.  We are not sure how close we are from reaching the minimum.
However, we believe that we sufficiently close to the minimum to allow us
to calculate $P$ and $R$ with a reasonable degree of reliability.  Figure~\ref{fig:pred} shows
a plot of SIDC sunspot number and the actual polar field data during the last
few years, indicating how we are obtaining $P$ and $R$.

We calculate 
the polar field $P$ near the minimum preceding cycle 25 by using polar field data from Wilcox Solar Observatory. We have taken one year average of the polar field before the present epoch, which is expected to be close to the minimum (see the red marked line in polar field curve of Figure~\ref{fig:pred}), to use this polar field as a precursor for predicting the cycle 25. 
Although throughout the paper we have used the calibrated long-term polar flux dataset from Mount Wilson Observatory \citep{Munoz13b}, we now use Wilcox Solar Observatory polar field data here because it is the most reliable 
directly measured available polar field dataset. The long-term data of polar flux, which we had used for obtaining the 
new precursor formulas (Equations~(\ref{formulla1}) and (\ref{formulla2})), has three cycles overlapping with the WSO direct polar field data. Therefore, we have calibrated the MWO polar flux data with respect to the WSO polar field data. After implementing the calibration factor, the polar flux near the minimum preceding solar cycle 25 corresponding to the WSO data turns out to be ($1.648 \pm 0.147$) $\times 10^{22}$ Mx. We use this value of $P$ for predicting the solar cycle 25. Next, we need to calculate the decay rate $R$ just before the minimum preceding cycle 25. We have calculated it
as indicated by the slope in Figure~\ref{fig:pred} with red line. This is found to be 14.84. We plug these values of $P$ and $R$ into the new precursor formulae (Equations~(\ref{formulla1}) and (\ref{formulla2})) 
to get an estimate of the peak amplitude of cycle 25. Please note that we have tentatively assumed the present epoch to be the minimum (black dashed line in Figure~\ref{fig:pred}) for the purpose of calculating $P$ and $R$. Since the actual minimum may occur 1-2 years later, the re-calculated values of $P$ and $R$ at that time may be slightly different, leading to a slightly modified prediction. 

According the formula in Equation~(\ref{formulla1}) (with $\alpha=\beta=0.5$), the peak amplitude 
of the cycle 25 would be 138 and with the formula in Equation~(\ref{formulla2}) ($\alpha = \beta = 1$) the peak amplitude would be 143 of SIDC data. These calculations suggest that
the cycle 25 would be a moderate cycle.           

\begin{figure}[!htbp]
\begin{center}
\includegraphics[width = 0.45\textwidth]{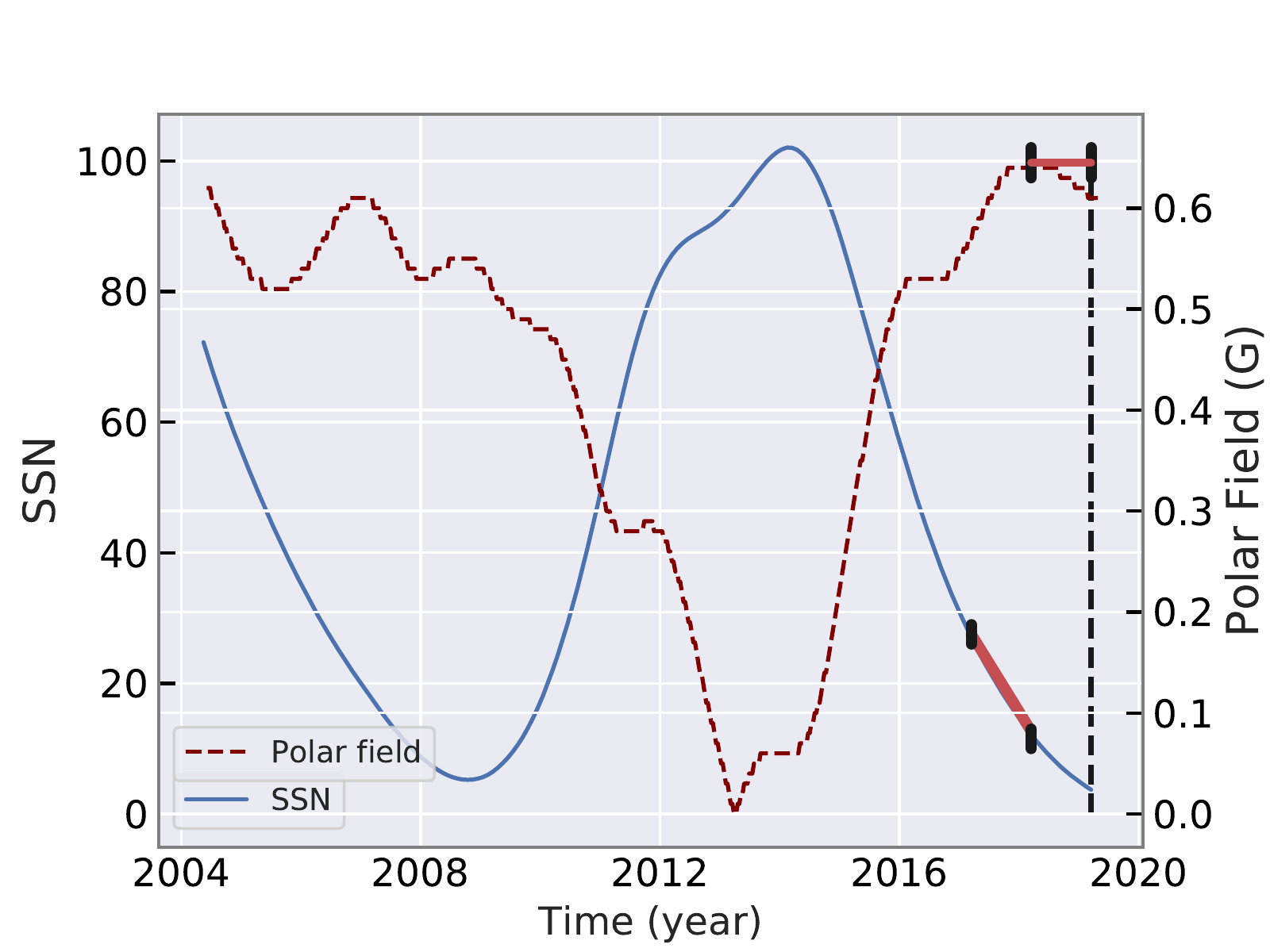}
\caption{\label{fig:pred} Sunspot number from SIDC and polar field from Wilcox Solar Observatory are shown in solid blue and dashed dark red lines respectively. The red marked line segments in sunspot number curve and polar field curve show the regions where we have calculated the decay rate ($R$) and polar field ($P$) respectively to predict the amplitude of the cycle 25. The black dashed vertical line shows the current epoch, which we tentatively assume to be the minimum to calculate $R$ and $P$.}
\end{center}
\end{figure}

\section{Conclusion}\label{sec:con}

Predicting a solar cycle before its onset is a challenge which interests both solar physicists and the general public.
Even a decade ago, it was a rather uncertain art. \citet{Pesnell08} combined all the predictions that were made for cycle~24
in Figure~1 of his paper.  It was clear that the various predictions covered virtually the entire range of all possible
values of the peak sunspot number. During the intervening years, our understanding of the physical basis for the solar
cycle prediction has deepened considerably.  The aim of the present paper has been to come up with a simple formula for predicting the
forthcoming cycle on the basis on this new understanding. 

We now believe that the irregularities of the solar cycle are produced primarily by two factors: fluctuations in the BL mechanism
and fluctuations in MC.  So, in order to predict a cycle, we need to include contributions from both these factors.  Since
the polar field $P$ at the beginning of the cycle provides the relevant information for fluctuations in the BL mechanism and
the decay rate $R$ at the end of the previous cycle provides the relevant information for fluctuations in MC, we have looked
for formulae combining $P$ and $R$ which have good correlation with the peak of the next cycle.  The formulae have to be
calibrated by using the data of the past cycles. The first bottleneck in this process is the lack of polar field data before
the 1970s.  We have pointed out three proxies for the polar field.  During much of the time, these three proxies give very
similar values for the polar field. However, there have been intervals during which some of the proxies diverged and we
do not have reliable information about the polar field in those intervals.  We saw that some of the data points in our
correlation plots with largest scatters corresponded to these intervals.
  
Since the polar field alone has often been used as a predictor for the
next cycle \citep{Svalgaard05,Schatten05}, we now come to the question 
whether it is really necessary to include $R$ in our formulae.
From the durations of past cycles, as shown in Figure~2 of \citet{KarakChou11},
it appears that there have been significant fluctuations in MC during the nineteenth
century, but there have not been very large fluctuations during cycles 15--22 
covering much of the twentieth century.  
This means that the polar field of the Sun alone would have been a reasonably good precursor for solar cycles during the twentieth century. Although we have no
idea at the present time as to what causes these fluctuations in MC, there is
no reason to expect that the nineteenth century was a very atypical era and we
are quite likely to enter similar eras of large fluctuations in MC in the 
future.  In such an era, using the polar field alone as a predictor for cycles
would probably be inadequate and we have to include the effect of varying MC,
as we have tried to do in this paper. We are forced to use data 
only for cycles 15--22 when some information about the polar field is available
from various proxies. On comparing Figure~2(a) of this paper with 
Figure~2(b) of \citet{HKBC15}, we find that most of the data points we are
considering now (except the points for cycles 19 and 22) have a rather 
narrow spread in the values of decay rate (horizontal axis) compared to what
we see in Figure~2(b) of \citet{HKBC15}. This certainly makes it difficult
to calibrate our formulae by incorporating the dependence of $R$ properly.
So, our formulae should be taken as provisional at the present time. 
Probably our formulae will get properly calibrated
only after we have an era of a few decades during which
the MC has large fluctuations (like the nineteenth century).

We may point out that there is a periodic variation of MC with the solar cycle 
\citep{komm1993, CD01, BA10, Gonzalez10, Hathaway11}, presumably due to
the Lorentz force of the dynamo-generated magnetic field \citep{HC17}.
There is also some evidence of a migratory pattern in MC variations with
migration of the activity belt, indicating flow towards this belt---at least
at the surface \citep{Snodgrass96, CD01, CS10, Howe18, Komm18, Lin18}. Presumably, the
amplitude of these inter-cycle variations would depend on the strength of the
cycle, but we have not considered these 
so far poorly understood variations in this work, limiting the
scope of our model by the inclusion of only a simple kind of nonlinearity 
in magnetic buoyancy (allowing the toroidal field to rise only when it is
stonger than $B_c$) to restrain the dynamo growth. So far there have not
been many studies of the nonlinear interaction between the MC and the
dynamo \citep{KarakChou12}.
Future studies should include the tilt-angle quenching \citep{KM17} and the nonlinear modulation due to active regions inflows \citep{Martin17}.

In summary, we have to say that the formulae proposed in this paper are of somewhat provisional nature
at the present time, since we have
very limited amount of past data to calibrate these formulae.  However, we believe that these formulae show the right
way for predicting future solar cycles.  
Based on the formulae we have arrived at, we have
presented our prediction for the upcoming cycle~25.
We expect that the formulae for predicting future cycles 
will get improved as solar astronomers 
get more data to calibrate
them in future and will eventually prove very powerful tools for predicting solar cycles before their onset. 

We thank two anonymous referees for their constructive comments which help to improve the manuscript. This work is partially supported by the JC Bose Fellowship
awarded to A.R.C. by the Department of Science and Technology, Government of India. 
G.H. is supported by the Zhuoyue Postdoctoral Fellowship, Beihang University, China. G.H. also acknowledges the support by the National Science Foundation of China (grant Nos. 11522325, 11873023, 41404136, and 11573038) and by the Fundamental Research Funds for the Central Universities of China.
\bibliography{myref}
\end{document}